\documentclass[11pt]{article}
\usepackage{latexsym}
\usepackage{graphicx}
\usepackage{dcolumn}
\usepackage{bm}
\usepackage{xcolor}
\input amssym.def
\input amssym.tex

\newcommand{\half}{{{\textstyle\frac{1}{2}}}}
\newcommand{\quarter}{{{\textstyle\frac{1}{4}}}}
\newcommand{\be}{\begin{equation}}
\newcommand{\ee}{\end{equation} }
\newcommand{\beqa}{\begin{eqnarray} }
\newcommand{\eeqa}{\end{eqnarray} }
\newcommand{\ba}{\begin{array}}
\newcommand{\ea}{\end{array}}

\newcommand\Tr{{\rm Tr}}

\newcommand\rd{{\rm d}}

\newcommand\cG{{\cal G}}

\newcommand\cL{{\cal L}}
\newcommand\cM{{\cal M}}

\newcommand\hd{n}

\newcommand\NaGo{{\rm \scriptscriptstyle{N.G.}}}
\newcommand\Poly{{{\rm \scriptscriptstyle Poly.}}}
\newcommand\New{{\scriptscriptstyle{\rm New}}}

\newcommand\NB{{\rm \scriptscriptstyle{N.B.}}}

\newcommand\passiv{{\rm \scriptscriptstyle{pssv.}}}
\newcommand\activ{{\rm \scriptscriptstyle{activ.}}}
\newcommand\aux{{\omega}}

\newcommand\dis{\displaystyle}

\def\tK{\tilde{K}}

\begin{document}
\begin{titlepage}
\title{\vskip -110pt 
\vskip 2cm   \mbox{Taking off the square root of  Nambu-Goto action and} obtaining    Filippov-Lie algebra   gauge theory  action~\\~}

\author{\sc Jeong-Hyuck Park${}^{\dagger}$ and   Corneliu Sochichiu${}^{\natural}$}
\date{}
\maketitle \vspace{-1.0cm}
\begin{center}
~~~\\
${}^{\dagger}$Department of Physics, Sogang University\\ Shinsu-dong, Mapo-gu, Seoul 121-742, Korea\\
\texttt{park@sogang.ac.kr}\\
~{}\\
${}^{\natural}$University College, Sungkyunkwan University\\Suwon 440-746, Korea\\
\texttt{sochichi@skku.edu}\\
~{}\\
~~~\\
~~~\\
\end{center}
\begin{abstract}
\vskip0.5cm
\noindent
We propose a novel prescription to take off the square root of  Nambu-Goto action for a  $p$-brane, 
which generalizes the Brink-Di Vecchia-Howe-Tucker or also known as Polyakov method.
With an arbitrary decomposition as $d+\hd=p+1$,   our resulting action is a modified  $d$-dimensional  
Polyakov action which is gauged and possesses  a Nambu $\hd$-bracket squared potential. 
We first spell out how the  $(p+1)$-dimensional diffeomorphism   is realized in the lower dimensional  action. 
Then we discuss  a  possible gauge fixing of it  to    a direct product of  $d$-dimensional diffeomorphism and  
$\hd$-dimensional volume preserving  diffeomorphism. We show that  the latter  naturally leads to a novel 
Filippov-Lie  $\hd$-algebra  based  gauge theory action in $d$-dimensions.
\end{abstract}
\vskip 20pt

{\small
\begin{flushleft}
~~~~~~~~\textit{Keywords}: Nambu-Goto action, Nambu-bracket, Filippov-Lie algebra\\~~\\~~\\
${}^{\natural}${On leave of absence from IAP AS 5 Academy Str. MD-2028 Chisinau, Moldova}
\end{flushleft}}

\thispagestyle{empty}
\end{titlepage}
\newpage


\section{Introduction}
A $p$-brane is a  spatially  extended object  
propagating   in a target spacetime. The number  $p$ counts   spatial dimensions of the brane such that 
$p=0,1,2,\cdots$ correspond to  point-like particle, string, membrane \textit{etc.}  The geodesic motion of a point particle 
\textit{i.e.~}${p=0}$ brane,  minimizes  the relativistic length of the trajectory in the target spacetime. 
Nambu-Goto action for a $p$-brane then generalizes this geometric significance: the induced worldvolume of the brane is to be minimized.

With an embedding of  ${({p+1})}$-dimensional worldvolume coordinates  into  $D$-dimensional target spacetime,
\be
X(\xi)~:~\xi^{m}~~\longrightarrow~X^{M}\,,
\ee
where $m=0,1,\cdots, p$ and $M=0,1,\cdots,{D-1}$, the Nambu-Goto action for a $p$-brane is~\cite{NambuGoto}
\be
S_{\NaGo}=\displaystyle{-\int}\rd^{p{+1}}\xi~\sqrt{-\det\cG_{mn}}\,.
\label{NGp}
\ee
Here  $\cG_{mn}$ is the induced metric onto the worldvolume such that the action  measures the relativistic worldvolume   of the $p$-brane in the target spacetime,
\be
\cG_{mn}:=\partial_{m}X^{M}\partial_{n}X^{N}G_{MN}(X)\,.
\label{inducedg}
\ee
For simplicity we set the brane tension to  unit.

Despite of its elegant  geometric significance,  Nambu-Goto action is hard to quantize due to the presence of a  highly nonlinear structure, 
the square root. An equivalent but far more convenient action is available, thanks to Deser-Zumino~\cite{Deser:1976rb}, 
Brink-Di Vecchia-Howe~\cite{Brink:1976sc}    and Howe-Tucker~\cite{Howe:1977hp},   by    introducing  an auxiliary worldvolume metric $h_{mn}$:
\be
\displaystyle{S_{\Poly}=-\half\displaystyle{\int}\rd^{p{+1}}\xi~
\sqrt{-h\,}\Big[\,h^{mn}\partial_{m}X^{M}
\partial_{n}X_{M}+1-p\,\Big]\,.}
\label{actionPoly}
\ee
This action  is often dubbed Polyakov action. Integrating out the auxiliary worldvolume metric
using its equation of motion, $h_{mn}\equiv\partial_{m}X^{M}\partial_{n}X_{M}$ for  $p\neq 1$ or
$h_{mn}\propto\partial_{m}X^{M}\partial_{n}X_{M}$ for  $p=1$, the Polyakov action reduces to the Nambu-Goto action 
$S_{\Poly}\!\!\equiv S_{\NaGo}$. Here and henceforth we denote the on-shell equality as well as  gauge fixings  by  
`$\equiv$'  and the defining equality by `$:=$'.  Both  Nambu-Goto and  Polyakov actions (\ref{NGp}), (\ref{actionPoly}) are manifestly invariant under   the  $(p+1)$-dimensional  worldvolume diffeomorphisms.

In the present paper we generalize the Brink-Di Vecchia-Howe-Tucker-Polyakov method and  construct an action whose  
characteristic  features are, compared to the Polyakov action, the appearance  of \textit{gauge covariant derivatives}  
and a \textit{Nambu bracket} squared potential.  After some gauge fixing we show that our action can be  identified as 
 a lower dimensional gauge theory action   based on      \textit{Filippov-Lie algebra}.

Previous works on related topics   include  the light-cone gauge fixed action of a $p$-brane \cite{Hoppe,Bergshoeff:1988hw}.\footnote{The appearance of a gauge connection from diffeomorphism invariance  is 
also well known in Kaluza-Klein theory, see \cite{Salam:1981xd} and references therein.  However,
the gauge field in  Kaluza-Klein theory originates from the spacetime metric which is dynamical in gravity,
while in our action the gauge field is introduced as a non-dynamical  auxiliary variable.}  
Taking the light-cone gauge means fixing 
 the light-cone variable    to a classical on-shell value. Hence  the light-cone gauge  action   
 describes only a sector of classically fixed light-cone momentum and breaks the full background  isometry.  In contrast, 
 our resulting action is  covariant and the full background isometry  survives.  
 Furthermore, the covariant derivative in our action takes a different form and is  based on Filippov-Lie algebra.

 In particular,  applying our result to the Nambu-Goto action for a five-brane we obtain a  Filippov three-algebra based gauge theory action in three dimensions. 
 As we will see, the precise form of  the gauge covariant derivative and the presence of the three-algebra squared potential are identical to the Bagger-Lambert-Gustavsson description of multiple M2-branes~\cite{BL,Gustavsson:2007vu}.\\


\section{General analysis}
Our prescription to generalize the Brink-Di Vecchia-Howe-Tucker-Polyakov method first starts with dividing  formally
the $p$-brane  worldvolume dimension into two parts,
\be
1+p=d+\hd\,,
\ee
which corresponds to the  decomposition  of the  worldvolume coordinates into two sets:
\be
\{\,\xi^{m}\,\}=\{\,\sigma^{\mu}\,,\,\varsigma^{i}\}\,,
\ee
where  $\mu=0,1,\cdots, d-1$ and $i=1,\cdots,\hd$. The decomposition here is \textit{ a priori}
arbitrary as for any positive  integers $d,\hd$. One natural  application of the splitting will be the case where $p$-brane is extended over both compact and non-compact directions: In this case  we reserve $\varsigma^{i}$ for compact directions  and $\sigma^{\mu}$ for non-compact directions including time.     ÊÊÊÊÊÊ

According to the splitting,     the induced metric (\ref{inducedg}) decomposes into the following  $d{\times d}$, $d{\times\hd}$ and $\hd{\times\hd}$ blocks $K,B,V$ defined by
\be
\ba{lll}
K_{\mu\nu}:=\cG_{\mu\nu}\,,~&~
B_{\mu i}:=\cG_{\mu i}\,,~&~
V_{ij}:=\cG_{ij}\,.
\ea
\ee
The first crucial step in our formalism is to express the  determinant of the   $(p+1){\times (p+1)}$ induced metric  as a product of two
determinants of the smaller $d\times d$ and $\hd\times\hd$  matrices:
\be
\ba{ll}
\det\cG_{mn}=\det\tK\det V\,,~~&~~\tK:=K-BV^{-1}B^{T}\,.
\ea
\ee
This  follows  from the following simple observation:
\be
{\scriptsize{
\det\!\!\left(\!\ba{cc} K &\!B \\{}&{}\\B^{T}&\!V\ea\right)\!=\det\!\!\left[\!\left(\ba{cc} 1 &\!-BV^{-1}\! \\{}&{}\\0&\!1\ea\right)\!\!\!
\left(\ba{cc} K &\!B \\{}&{}\\B^{T}&\!V\ea\right)\!\right]\!=\det\!\!\left(\ba{cc} \tK &\!0 \\{}&{}\\B^{T}&\!V\ea\right).}}
\ee
The resulting Nambu-Goto action (\ref{NGp})
\be
\displaystyle{S_{\NaGo}=-\int\rd^{p+1}\xi~\sqrt{-\det\tK\det V}\,,}
\label{pNG}
\ee
can be now reformulated in a square root free form, if we   introduce an auxiliary variable $\rho$:
\be
\displaystyle{\int\rd^{p+1}\xi\left(\rho\det\tK -\quarter\rho^{-1}\det V\right)\,.}
\label{tKVaction}
\ee
To proceed further, we introduce a $d\times d$ auxiliary matrix  $\varphi_{\mu\nu}$ and  apply  the  Brink-Di Vecchia-Howe-Tucker-Polyakov method to the determinant of $\tK_{\mu\nu}$ in order to have our semi-final action:
\be
\displaystyle{\int\rd^{p{+1}}\xi\,\left[
\rho\det\!\varphi\left(\varphi^{\mu\nu}\tK_{\mu\nu}+1-d\right) -\quarter\rho^{-1}\det V\right]\,.}
\label{varphiaction}
\ee
At this point it is convenient to reparameterize    the auxiliary variables $\rho$,  $\varphi_{\mu\nu}$ by a new auxiliary scalar $\aux$ and a  $d$-dimensional `worldvolume metric' $h_{\mu\nu}$ as
\be
\ba{ll}
\aux^{-1}:=(-\rho^{2}\det\varphi)^{\frac{1}{d-2}}\,, ~& h_{\mu\nu}:=(-\rho^{2}\det\varphi)^{\frac{1}{d-2}}\varphi_{\mu\nu}\,.
\ea
\label{sing}
\ee
Now we are ready to spell  our novel action, which we propose in order to reformulate  the  Nambu-Goto action for  $p$-brane:
\be
\ba{l}
\dis{S_{\New}=\int\rd^{d}\sigma~\Tr\left(\sqrt{-h\,}\cL_{\New}\right)\,,}~~~~~~~\dis{\Tr:=\int\rd^{\hd}\varsigma\,,}\\
{}\\
\cL_{\New}=-h^{\mu\nu}D_{\mu}X^{M}D_{\nu}X_{M}
-\textstyle{\frac{1}{4}}\aux^{d-1}\det V+(d-1)\aux\,.
\ea
\label{actionNew}
\ee
In addition to $\aux$ and $h_{\mu\nu}$, here we introduced  one more  auxiliary  field
$A_{\mu}^{~i}$ which defines the  `covariant derivative':
\be
D_{\mu}X^{M}:=\partial_{\mu}X^{M}-A_{\mu}^{~i}\partial_{i}X^{M}\,.
\ee
The corresponding field strength reads
\be
F_{\mu\nu}^{~i}=\partial_{\mu}A_{\nu}^{~i}-\partial_{\nu}A_{\mu}^{~i}-A_{\mu}^{~j}\partial_{j}A_{\nu}^{~i}+
A_{\nu}^{~j}\partial_{j}A_{\mu}^{~i}\,.
\ee
In terms of the Nambu $\hd$-bracket  which is defined by \cite{Nambu:1973qe}
\be
\{Y_{1},Y_{2},\cdots,Y_{\hd}\}_{\NB}
:=\epsilon^{i_{1}i_{2}\cdots i_{{\hd}}}\partial_{i_{1}}Y_{1}\partial_{i_{2}}Y_{2}\cdot\cdot\partial_{i_{\hd}}Y_{\hd}\,,
\ee
the `potential' $\det V$ takes   the form:\footnote{For the curved target spacetime manifold having the metric $G_{MN}(X)$,  one should bear in mind that
$D_{\mu}X_{N}=D_{\mu}X^{M}G_{MN}$ and  $\{X_{N_{1}},X_{N_{2}},\cdots,X_{N_{\hd}}\}_{\NB}=
\{X^{M_{1}},X^{M_{2}},\cdots,X^{M_{\hd}}\}_{\NB}G_{M_{1}N_{1}}G_{M_{2}N_{2}}\cdots G_{M_{\hd}N_{\hd}}$.}
\be
\det V\!=\textstyle{\frac{1}{\hd!}}\{X^{M_{1}},X^{M_{2}},\cdot\cdot,X^{M_{\hd}}\}_{\NB}
\{X_{M_{1}},X_{M_{2}},\cdot\cdot,X_{M_{\hd}}\}_{\NB}\,.
\label{potentialNB}
\ee
In the above  $\epsilon^{i_{1}i_{2}\cdots i_{{\hd}}}$ is   the totally anti-symmetric $\hd$-dimensional  tensor of the normalization  $\epsilon^{12\cdots{\hd}}=1$.

The auxiliary variables assume the on-shell values:
\be
\ba{lll}
A_{\mu}^{~i}\equiv(BV^{-1})_{\mu}{}^{i}\,,~&\aux^{2-d}\equiv \quarter\det V\,,~&
h_{\mu\nu}\equiv\aux^{-1}\tK_{\mu\nu}\,.
\ea
\label{onshellaux}
\ee
Plugging these into the action~(\ref{actionNew}),  we recover the Nambu-Goto action~(\ref{pNG}), $S_{\New}\equiv S_{\NaGo}$.  In particular, we have the following on-shell relations,
\be
\ba{ll}
\partial_{i}X^{M}D_{\mu}X_{M}\equiv 0\,,~~&~~
D_{\mu}X^{M}D_{\nu}X_{M}\equiv\tK_{\mu\nu}\,.
\ea
\ee
The former is nothing  but the Euler-Lagrangian equation for $A_{\mu}^{~i}$ which is solved by $A\equiv BV^{-1}$ and prescribes that  $D_{\mu}X^{M}$ should be orthogonal to $\partial_{i}X^{M}$ on-shell. The latter holds since $P^{M}_{~N}:=\delta^{M}_{~N}-\partial_{i}X^{M}(V^{-1})^{ij}\partial_{j}X_{N}$ is a  projector satisfying $P^{2}=P$. Note also  that $~P^{M}_{~N}\partial_{i}X^{N}=0$ and $~P^{M}_{~N}\partial_{\mu}X^{N}\equiv D_{\mu}X^{M}$. \\

 Although not manifest,   our  novel action  (\ref{actionNew}) enjoys   the full  $(p+1)$-dimensional  diffeomorphism symmetry  like the Nambu-Goto action (\ref{NGp}),  irrespective of the arbitrary splitting of the worldvolume coordinates: Under an arbitrary  infinitesimal   coordinate transformation $\delta\xi^{m}=-\upsilon^{m}$ or $\delta\partial_{m}=\partial_{m}\upsilon^{n}\partial_{n}$, all the fields transform as
\be
\ba{l}
\delta X^{M}=0\,,\\{}\\
\delta A_{\mu}^{~i}=D_{\mu}\upsilon^{\nu}A_{\nu}^{~i}+D_{\mu}\upsilon^{i}
+\textstyle{\frac{1}{4}}\aux^{d-1}\det V h_{\mu\nu}\partial_{j}\upsilon^{\nu}V^{-1 ji}\,,\\{}\\
\delta\aux=-\textstyle{\frac{2}{d-2}}\aux(\partial_{i}\upsilon^{\lambda}A_{\lambda}^{~i}
+\partial_{i}\upsilon^{i})\,,\\{}\\
\delta h_{\mu\nu}=D_{\mu}\upsilon^{\lambda}h_{\lambda\nu}+D_{\nu}\upsilon^{\lambda}h_{\mu\lambda}
+\textstyle{\frac{2}{d-2}}(\partial_{i}\upsilon^{\lambda}A_{\lambda}^{~i}+\partial_{i}\upsilon^{i})h_{\mu\nu}\,.
\ea
\label{diffgen}
\ee
Note that this transformation rule  is consistent with the on-shell relations (\ref{onshellaux}), and further that
we assume  the `active' form of the diffeomorphism. The dual `passive' diffeomorphism which is directly relevant to the Noether symmetry is given by $\delta_{\passiv}\partial_{m}=0$ and $\delta_{\passiv}{\Phi}=
 \delta_{\activ}{\Phi} + v^{m}\partial_{m}{\Phi}$  for each field $\Phi$.  \\

Apparently from (\ref{sing}), the above formalism is singular if ${d=2}$,  essentially  due to the Weyl invariance in two dimensions.  In this case, we return to (\ref{varphiaction}), let $h_{\mu\nu}:=\varphi_{\mu\nu}$ and introduce a dilaton $e^{-\phi}:=\rho\sqrt{- h}$. The proposed action for ${d=2}$ case becomes, rather than (\ref{actionNew}):
\be
\ba{l}
\dis{S^{d=2}_{\New}=\int\rd^{2}\sigma~\Tr\left(\sqrt{-h\,}\cL^{d=2}_{\New}\right)\,,}
~~~~~~~\dis{\Tr:=\int\rd^{\hd}\varsigma\,,}\\
{}\\
\cL^{d=2}_{\New}=-e^{-\phi}h^{\mu\nu}D_{\mu}X^{M}D_{\nu}X_{M}
-\textstyle{\frac{1}{4}}e^{\phi}\det V+e^{-\phi}\,.
\ea
\label{actionNew2}
\ee
Like (\ref{onshellaux}) the auxiliary variables assume the following on-shell values:
\be
\ba{lll}
A_{\mu}^{~i}\equiv(BV^{-1})_{\mu}{}^{i}\,,~&e^{-2\phi}\equiv \quarter\det V\,,~&
h_{\mu\nu}\equiv\tK_{\mu\nu}\,.
\ea
\label{onshellaux2}
\ee
Plugging these into the action~(\ref{actionNew2}) we recover the Nambu-Goto action~(\ref{NGp}) again.
The full  ${(p+1)}$-dimensional  diffeomorphism symmetry has the following   two-dimensional realization:
\be
\ba{l}
\delta X^{M}=0\,,\\{}\\
\delta A_{\mu}^{~i}=D_{\mu}\upsilon^{\nu}A_{\nu}^{~i}+D_{\mu}\upsilon^{i}
+\textstyle{\frac{1}{4}}e^{2\phi}\det V h_{\mu\nu}\partial_{j}\upsilon^{\nu}V^{-1 ji}\,,\\{}\\
\delta\phi=-\partial_{i}\upsilon^{\lambda}A_{\lambda}^{~i}-\partial_{i}\upsilon^{i}\,,\\{}\\
\delta h_{\mu\nu}=D_{\mu}\upsilon^{\lambda}h_{\lambda\nu}+D_{\nu}\upsilon^{\lambda}h_{\mu\lambda}\,.
\ea
\label{diffgen2}
\ee
~\\

Although the action (\ref{actionNew}) is still valid except  $d=2$, the case of  $d=1$ is special: the auxiliary scalar
$\aux$ drops from the action as well as from the diffeomorphism transformations. In other words,  when $d=1$  we need only two types of auxiliary fields to take off the square root of the Nambu-Goto action of a $p$-brane: an einbein $e$ and a  gauge field $A_{\tau}^{~i}$,  $i=1,2,\cdots,p$.   With a worldline parameter $\tau$,
the action (\ref{actionNew}) reduces to
\be
\dis{S_{\New}^{d=1}=
\int\rd\tau~\Tr\Big(e^{-1}D_{\tau}X^{M}D_{\tau}X_{M}
-\textstyle{\frac{1}{4}}e\det V\Big)\,.}
\label{actionNew1}
\ee
The on-shell values of  the auxiliary fields are  then:
\be
\ba{ll}
A_{\tau}^{~i}\equiv(BV^{-1})_{\tau}{}^{i}\,,~&~
e \equiv 2\sqrt{-D_{\tau}X^{M}D_{\tau}X_{M}/\det V\,}\,.
\ea
\label{onshellaux1}
\ee
In this case of ${d=1}$   the full ${(p+1)}$-dimensional  diffeomorphism takes the following form:
\be
\ba{l}
\delta X^{M}=0\,,\\{}\\
\delta A_{\tau}^{~i}=D_{\tau}\upsilon^{\tau}A_{\tau}^{~i}+D_{\tau}\upsilon^{i}
-\textstyle{\frac{1}{4}}e^{2}\det V \partial_{j}\upsilon^{\tau}V^{-1 ji}\,,\\{}\\
\delta e=e\left(D_{\tau}\upsilon^{\tau}-A_{\tau}^{~i}\partial_{i}\upsilon^{\tau}-\partial_{i}\upsilon^{i}\right)\,.
\ea
\label{diffgen1}
\ee
~\\
\section{Gauge fixing to  Filippov-Lie  $\hd$-algebra}
Although our resulting actions for  a $p$-brane, (\ref{actionNew}) for ${d\geq 3}$, (\ref{actionNew2}) for ${d=2}$ and (\ref{actionNew1})  
for ${d=1}$ are written  in the form of a  $d$-dimensional gauge theory with $d$ being less that ${p+1}$,  
they are invariant under the full ${(p+1)}$-dimensional  diffeomorphism. They are still  identified as      
$(p+1)$-dimensional models. In order to be identified as  genuine lower dimensional gauge theories, it is necessary to break the full  
${(p+1)}$-dimensional  diffeomorphism to  a direct product of the  $d$-dimensional diffeomorphism and the $\hd$-dimensional 
volume preserving  diffeomorphism. The latter then corresponds  to a local gauge symmetry of the $d$-dimensional action.  
In fact, for each value of $d$ we can impose  a pair of  gauge fixing conditions:\footnote{
When $d=1$, besides (\ref{gf1}), it is also possible to set  $e\equiv 2$ and $A_{\tau}^{i}\equiv 0$ for all $i=1,2,\cdots,p$,  
utilizing  the full $(p+1)$-dimensional worldvolume diffeomorphism.   
Then the case of $p=1$ coincides with  the well known conformally gauge fixed  Polyakov string action.
We thank Kanghoon Lee for pointing out this~\cite{Kanghoon}.}
\begin{itemize}
\item For $d\geq 3$,
\be
\ba{ll}
\partial_{i}A_{\mu}^{~i}\equiv 0\,,~~~~&~~~~\aux\equiv 1\,.
\ea
\label{gf3}
\ee
The unbroken local symmetry is then the direct product of the  $d$-dimensional diffeomorphism and the  $\hd$-dimensional volume preserving  gauge symmetry, generated by the infinitesimal transformations satisfying   $\partial_{i}\upsilon^{\mu}=0$ and  $\partial_{j}\upsilon^{j}=0$.
\item For $d=2$,
\be
\ba{ll}
\partial_{i}A_{\mu}^{~i}\equiv 0\,,~~~~&~~~~\phi\equiv 0\,.
\ea
\label{gf2}
\ee
The unbroken local symmetry is  the direct product of the  two-dimensional diffeomorphism and the  $(p-1)$-dimensional volume preserving  gauge symmetry.
\item For $d=1$,
\be
\ba{ll}
\partial_{i}A_{\tau}^{~i}\equiv 0\,,~~~~&~~~~e\equiv 2\,.
\ea
\label{gf1}
\ee
As we fix the einbein, the unbroken local gauge symmetry is given by  the  $p$-dimensional volume preserving diffeomorphism only.
\end{itemize}
In each case, from (\ref{diffgen}), (\ref{diffgen2}),  (\ref{diffgen1}), the former $d$-number of  conditions can be essentially  achieved by  diffeomorphism with the $d$-number of $\upsilon^{\mu}$ generators satisfying   $\partial_{i}\upsilon^{\mu}\neq 0$, while the latter single condition can be met by   $\partial_{i}\upsilon^{i}\neq 0$.\\

The divergence free condition $\partial_{i}A_{\mu}^{~i}\equiv 0$ must be imposed  once we demand
 the covariant derivative $D_{\mu}=\partial_{\mu}-A_{\mu}^{~i}\partial_{i}$ to be an anti-Hermitian differential operator,
 allowing the usual  \textit{integration by parts}.  Furthermore,   the volume preserving diffeomorphism generators also satisfy the divergence free condition  $\partial_{i}\upsilon^{i}=0$.  That is to say, as usual,  the gauge connection assumes   the same  ``Lie algebra" value as the volume preserving  gauge symmetry generators. \\

Now it is crucial to note that the volume preserving  gauge symmetry generator    as well as the covariant derivative can be represented by  the Nambu $\hd$-bracket:\footnote{From the Poincare lemma  the divergence free  volume preserving  generator is given by $\upsilon^{i}\partial_{i}=\epsilon^{i_{1}i_{2}\cdots i_{\hd}}\partial_{i_{1}}\hat{\upsilon}_{i_{2}\cdots i_{\hd-1}}\partial_{i_{\hd}}$ which can be further organized to take  the form  (\ref{upDNB}).}
With the functional basis  $T^{a}(\varsigma)$, $a=1,2,3,\cdots$ for the $\hd$-dimensional manifold  we have
\be
\ba{l}
\upsilon^{i}\partial_{i}=\upsilon_{a_{1}a_{2}\cdots a_{\hd-1}}\{T^{a_{1}},T^{a_{2}},\cdots,T^{a_{\hd-1}},~~~\}_{\NB}\,,\\~\\
D_{\mu}=\partial_{\mu}-A_{\mu a_{1}a_{2}\cdots a_{\hd-1}}\{T^{a_{1}},T^{a_{2}},\cdots,T^{a_{\hd-1}},~~~\}_{\NB}\,.
\label{upDNB}
\ea
\ee
Note that here $\upsilon_{a_{1}a_{2}\cdots a_{\hd-1}}$ and $A_{\mu a_{1}a_{2}\cdots a_{\hd-1}}$ are $d$-dimensional fields, being independent of the $\varsigma^{i}$ coordinates. Further the $\hd$-dimensional manifold   is assumed  to be compact.\\

As is well known (see \textit{e.g.~}\cite{Takhtajan:1993vr}), Nambu $\hd$-bracket provides an explicit  realization  of  infinite dimensional 
Filippov-Lie  $\hd$-algebra~\cite{n-Lie}  defined by   $\hd$-bracket satisfying the totally anti-symmetric property:
\be
[X_{1},\cdots,X_{i},\cdots,X_{j},\cdots,X_{\hd\,}]=-[X_{1},\cdots,X_{j},\cdots,X_{i},\cdots,X_{\hd\,}]\,,
\label{antisym}
\ee
and the Leibniz rule, also known as a fundamental identity:
\be
\left[X_{1},\cdots,X_{{\hd-1}},[Y_{1},\cdots,Y_{\hd\,}]\right]=\sum_{j=1}^{\hd}~
\left[Y_{1},\cdots,[X_{1},\cdots,X_{{\hd-1}},Y_{j\,}],\cdots,Y_{\hd\,}\right]\,.
\label{Leibniz}
\ee
In the Nambu-bracket representation of  a  Filippov-Lie algebra, we may employ the structure constant through
\be
\{T^{a_{1}},T^{a_{2}},\cdots,T^{a_{\hd}}\}_{\NB}=f^{a_{1}a_{2}\cdots a_{\hd}}{}_{b}T^{b}\,.
\label{structureconstant}
\ee
The structure constant  is then totally anti-symmetric for the upper indices and  satisfies from the Leibniz rule (\ref{Leibniz}):
\be
f^{a_{1}a_{2}\cdots a_{\hd}}{}_{c}f^{b_{1}b_{2}\cdots b_{\hd}}{}_{a_{\hd}}=\sum_{j=1}^{\hd}~
f^{a_{1}a_{2}\cdots a_{{\hd{-}1}}b_{j}}{}_{e}\,f^{b_{1}\cdots b_{{j{-}1}}eb_{{j{+}1}}\cdots b_{\hd}}{}_{c}\,.
\label{Leibnizf}
\ee
Now from (\ref{upDNB}) and (\ref{structureconstant}), expanding the dynamical variables by the  functional basis $X^{M}(\sigma,\varsigma)=X^{M}_{a}(\sigma)T^{a}(\varsigma)$, the covariant derivative can be rewritten as
\be
\ba{ll}
D_{\mu}X^{M}=(D_{\mu}X^{M})_{a}T^{a}\,,~~~~&~~~~(D_{\mu}X^{M})_{a}=\partial_{\mu}X^{M}_{a}-X^{M}_{b}\tilde{A}_{\mu}^{b}{}_{a}\,,
\ea
\label{covDBLG1}
\ee
where we set
\be
\tilde{A}_{\mu}^{b}{}_{a}:=A_{\mu c_{1}c_{2}\cdots c_{\hd-1}}f^{c_{1}c_{2}\cdots c_{\hd-1}b}{}_{a}\,.
\label{covDBLG2}
\ee
In this way, after the gauge fixings,  our final actions (\ref{actionNew}), (\ref{actionNew2}), (\ref{actionNew1}) reduce  to genuine  lower dimensional Filippov-Lie  
$\hd$-algebra  based gauge theory actions, where the potential is given by the $\hd$-Lie bracket squared (\ref{potentialNB}) and the covariant derivative is given by (\ref{covDBLG1}).   Furthermore, at this point, we may generalize the actions to assume  an  arbitrary (finite or infinite dimensional)  
 Filippov-Lie  $\hd$-algebra as a gauge symmetry.  With $\tilde{v}^{b}{}_{a}:=v_{c_{1}c_{2}\cdots c_{\hd-1}}f^{c_{1}c_{2}\cdots c_{\hd-1}b}{}_{a}$, from
the passive transformation of (\ref{diffgen}) and the expression (\ref{upDNB}), the Filippov-Lie  $\hd$-algebra  based gauge transformation is  given by
\be
\ba{l}
\delta X^{M}_{a}=X^{M}_{b}\tilde{v}^{b}{}_{a}\,,\\
~\\
\delta A_{\mu a_{1}a_{2}\cdots a_{\hd-1}}=\partial_{\mu}v_{a_{1}a_{2}\cdots a_{\hd-1}}+
(-1)^{\hd}(\hd-1)A_{\mu c[a_{1}a_{2}\cdots a_{\hd-2}}\tilde{v}^{c}{}_{a_{\hd-1}]}\,,
\ea
\ee
of which the latter induces, from (\ref{Leibnizf}),
\be
\delta \tilde{A}_{\mu}^{b}{}_{a}=\partial_{\mu}\tilde{v}^{b}{}_{a}-\tilde{v}^{b}{}_{c}\tilde{A}_{\mu}^{c}{}_{a}+
\tilde{A}_{\mu}^{b}{}_{c}\tilde{v}^{c}{}_{a}\,.
\ee

Especially, taking $\hd=3$, equations (\ref{covDBLG1}) and (\ref{covDBLG2}) precisely coincide with the definition of the covariant derivative in the Bagger-Lambert-Gustavsson description of multiple M2-branes \textit{via} Filippov three-algebra gauge interaction~\cite{BL,Gustavsson:2007vu}.\\

\section{Comments}
Filippov-Lie  $\hd$-algebra  is normally equipped with a bi-linear inner product.
This might be a potential  problem whilst  identifying  our final actions  (\ref{actionNew}), (\ref{actionNew2}), (\ref{actionNew1}) 
after the  gauge fixing (\ref{gf3}), (\ref{gf2}),  (\ref{gf1})  as a  $d$-dimensional gauge theory  based on a genuine 
Filippov-Lie  $\hd$-algebra, since the actions are not generically quadratic.    For example the kinetic term reads
\[
\sqrt{-h\,}h^{\mu\nu}D_{\mu}X^{M}D_{\nu}X_{M}\,.
\]
Again the Nambu-bracket provides a solution by simply generalizing the bi-linear inner product to multi-linear inner products or the ``trace" \cite{Lee:2009ue}:
\be
\Tr\left(T^{a}T^{b}\cdots T^{c}\right)=
\displaystyle{\int}{\rm d}^{\hd}\varsigma\,T^{a}T^{b}\cdots T^{c}\,,
\label{TRACE}
\ee
which is invariant under the  Filippov-Lie  $\hd$-algebra  gauge transformation: For arbitrary $m=1,2,3,\cdots$,
\be
\sum_{k=1}^{m}~
\Tr\Big( Y_{1},Y_{2},\cdots,Y_{k-1},[X_{1},
\cdots,X_{{\hd-1}},Y_{k}],Y_{{k+1}},\cdots\,Y_{m}\Big)=0\,,
\label{invbra2}
\ee
or equivalently
\be
\sum_{k=1}^{m}~
f^{a_{1}a_{2}\cdots a_{\hd{-}1}b_{k}}{}_{c}\Tr\Big(T^{b_{1}}T^{b_{2}}\cdots  T^{b_{k{-}1}}  T^{c} T^{b_{k{+}1}}\cdots  T^{b_{m}}\Big)=0\,.
\label{invbra2p}
\ee
~\\

Our work manifests the general phenomenon, commonly known as Myers effect~\cite{Myers:1999ps},  that non-Abelian structure of lower dimensional gauge theories can capture the description of a higher dimensional brane:\footnote{Since the trace (\ref{TRACE}) is invariant under any permutation of its arguments, the word `non-Abelian' might be improper.  More relevant structure appears to be the
Filippov-Lie  $\hd$-algebra.} a single $p$-brane can be described not  only by a $({p+1})$-dimensional Polyakov action but also by a gauged $d$-dimensional
Polyakov action based on Filippov-Lie  $\hd$-algebra with $p+1=d+\hd$. Since the functional basis of the  $\hd$-dimensional manifold is  
infinite dimensional, the corresponding gauge group based on the  Filippov-Lie  $\hd$-algebra 
is  \textit{a priori}  infinite dimensional. However, we emphasize that our final action admits 
a simple generalization taking   \textit{any} Filippov-Lie  algebra as a gauge symmetry.\footnote{For the discussion on the uniqueness of   
finite dimensional Filippov-Lie algebra see 
\cite{FigueroaO'Farrill:2002xg,Papadopoulos:2008sk,Gauntlett:2008uf,Papadopoulos:2008gh}.} 

 If we turn off the Filippov-Lie  $\hd$-algebra  gauge interaction,   
 our $d$-dimensional action  corresponds simply to a Polyakov action for $({d-1})$-brane. This suggests  the following  physical picture behind our formalism:  {the description of a single $p$-brane as a condensation of infinitely many lower dimensional branes through  
Filippov-Lie algebra gauge interactions}.

In particular, the  action (\ref{actionNew1}) provides a description of a $p$-brane \textit{via} infinitely many interacting relativistic  point-particles (see \cite{Yang:1998qd} for a related earlier work). Especially if we apply our formalism to an M2-brane in eleven dimensions   we obtain with the choice of $d=1$,
\be
\dis{S_{\rm M2}=\int\rd\tau~\Tr\Big(e^{-1}D_{\tau}X^{M}D_{\tau}X_{M}
-\textstyle{\frac{1}{8}}e[X^{M},X^{N}][X_{M},X_{N}]\Big)\,.}
\label{actionNewM2}
\ee
Since there are eleven scalars as $M=0,1,2,\cdots,10$, this action corresponds to a covariant version of
the $\cM$-theory matrix model~\cite{Banks:1996vh} (see also \cite{Sochichiu:2005ex}), without the light-cone gauge fixing.

Furthermore, in the case of $p=5$ and $d=\hd=3$, our results have common features with  the  Bagger-Lambert-Gustavsson description of multiple M2-branes~\cite{BL,Gustavsson:2007vu}:  
Filippov three-algebra naturally arises, the definition of the covariant derivative precisely  coincides and the potential is given by the three-bracket squared. This supports the idea that the Bagger-Lambert-Gustavsson action with  infinite dimensional gauge group may  describe a M5-brane as a condensation of infinitely many interacting  M2-branes, as explored in 
\cite{Ho,Krishnan:2008zm,Gomis:2008uv,Benvenuti:2008bt,Ho:2008ei,Jeon:2008bx,Bandos:2008fr,Bandos:2008df}.


\newpage

\noindent\textbf{Acknowledgements}\\
We wish to thank Xavier Bekaert,  Kanghoon Lee, Dmitri Sorokin for useful   comments and  especially  Choonkyu Lee for encouraging us to look for the full  diffeomorphism. The work is in part supported by   the Center for Quantum Spacetime of Sogang University with grant number R11 - 2005 - 021, and also   by the Korea Science and Engineering Foundation grant funded by the Korea government (R01-2007-000-20062-0).
\newpage



\end{document}